\documentclass[a4paper,12pt]{article}
\usepackage{graphicx,amsmath,amsfonts,amssymb,cite}
\newcommand{\be}{\begin{equation}}
\newcommand{\ee}{\end{equation}}
\newcommand{\ba}{\begin{eqnarray}}
\newcommand{\ea}{\end{eqnarray}}

\begin{document}

\begin{center}
{\Large\bf Hydrogen like classification for light nonstrange mesons}
\end{center}

\begin{center}
{\bf S. S. Afonin\footnote{On leave of absence from V. A. Fock Department of
Theoretical Physics, St. Petersburg State
University, 1 ul. Ulyanovskaya, 198504 St. Petersburg, Russia.}}
\end{center}

\begin{center}
{University of Bochum, Department of Physics and Astronomy,
Theoretical Physics II, 150 Universit\"{a}tsstrasse, 44780 Bochum,
Germany}
\end{center}

\begin{abstract}
The recent experimental results on the spectrum of highly excited
light nonstrange mesons are known to reveal a high degree of
degeneracy among different groups of states. We revise some
suggestions about the nature of the phenomenon and put the
relevant ideas into the final shape. The full group of approximate
mass degeneracies is argued to be $SU(2)_f\times I\times O(4)$,
where $I$ is the degeneracy of isosinglets and isotriplets and
$O(4)$ is the degeneracy group of the relativistic hydrogen atom.
We discuss the dynamical origin and consequences of considered
symmetry with a special emphasis on distinctions of this symmetry
from the so-called chiral symmetry restoration scenario.
\end{abstract}


\section{Introduction}

The discovery of approximate symmetries in the hadron spectrum
played an important role in establishing the structure of hadrons
and of underlying strong interactions. The observation of many new
resonances in recent years raised a renewed interest in the
spectral degeneracies. Broadly speaking, the problem can be framed
as follows: If a set of hadrons reveals a clear-cut clustering
near certain values of mass, what symmetry is responsible for the
observed pattern of approximate mass degeneracy and what are the
physical reasons for this symmetry? Needless to say, the correct
answer to this question can help considerably in unveiling the
underlying universal physics to the first approximation, the next
step would be the understanding of the sign and magnitude of fine
splittings inside degenerate multiplets, but those phenomena are
usually more involved and strongly channel-dependent (e.g., the
masses of resonances can be moved seriously by the threshold
effects).

A remarkable recent example of such a clustering is provided by
the spectrum of unflavored mesons, see~\cite{mpla} for a review.
The effect is certainly seen for the well confirmed states from
the Particle Data Group (PDG)~\cite{pdg}. A clear-cut cluster
structure of the spectrum of light nonstrange mesons was
convincingly confirmed by the Crystal Barrel experiment on
$\bar{p}p$ annihilation in flight~\cite{bugg,epja} which ran at
the antiproton storage ring LEAR at CERN. In brief, all observed
unflavored mesons above the chiral symmetry breaking (CSB) scale
in QCD, approximately 1~GeV, cluster into fairly narrow mass
ranges with the "centers of gravity" situated near 1340, 1700,
2000, and 2260~MeV. The corresponding spectrum is populated mainly
by the radial and orbital excitations of some ground states whose
masses lie below the CSB scale. The close values of masses in
these "towers" of resonances imply that the states inside each
cluster are related by some $\mathcal{X}$-symmetry of unknown
nature~\cite{mpla}.

The purpose of this paper is to develope a possible candidate
for the $\mathcal{X}$-symmetry. We will argue that the full
symmetry governing the approximate mass degeneracies in the light
nonstrange mesons seems to be
$\mathcal{X}=SU(2)_f\times I\times O(4)$,
where $I$ means the degeneracy of isosinglets and
isotriplets emerging due to the Zweig rule and $O(4)$ is the
hydrogen like degeneracy of energy levels.

An immediate question which likely rises here is how the
suggested $O(4)$ symmetry is related to QCD, i.e. how it can be
understood from the first principles? We do not know a convincing
answer to this question. We remind, however, that numerous
phenomena and symmetries emerging in the solid state physics
originate from Quantum Electrodynamics, on the other hand they are
not seen on the level of QED Lagrangian and hardy ever can be
derived from the underlying fundamental theory. Even in such a
simple system as the classical hydrogen atom, the $SO(4)$
symmetry of energy levels appears, which hardly can be envisaged
starting from the QED Lagrangian because it is a {\it dynamical}
symmetry reflecting the internal structure of the system, it seems
to have nothing to do with the approximate symmetries of the QED
Lagrangian. Needless to say that QCD is much more complex theory
and in the hadron world we can encounter manifestations of similar
dynamical symmetries, thus it is not excluded that searching for the
complete explanation of observed spectral symmetries in hadrons
having at hand the QCD Lagrangian only,
one is staying on a false way.

To clarify the point let us consider a simple example. The spin
$J$ and mass $m$ are known to be two independent Casimir
invariants of the Poincar\'{e} group. Hence, if there is a functional
dependence between $J$ and $m$, some higher symmetry takes place.
Consider now a classical object of size $r$ rotating at constant
velocity, its angular momentum is $J\sim mr$. On the other hand,
$n$-dimensional object of constant density has mass $m\sim r^n$,
i.e. $r\sim m^{1/n}$. Thus,
\be
\label{re}
J\sim m^{1+1/n}.
\ee
Quantum theory tells us that properties of any quantum system
approach to its classical ones if the quantum numbers defining the
stationary states of this system are large enough~\cite{landau}.
For this reason the highly excited hadrons are inevitably
quasiclassical objects, i.e. the classical arguments can be
applied for them as the first approximation. The functional
dependence $m(J)$ is an inherent feature of the Regge theory,
experimentally the Regge trajectories are approximately linear on
$(J,m^2)$ plane, at least for highly excited hadrons.
Consequently, to the extent that the Regge trajectories are
linear, the excited hadrons can be viewed quasiclassically as
one-dimensional objects of constant density according to
Eq.~\eqref{re}. Thus, one arrives at a nice agreement of very
general arguments and the real-life phenomenology. What must be
emphasized here is that relation~\eqref{re} is dynamical, we need
not any particular Lagrangian to obtain it --- the role of
interactions is to create the rotating system, the ensuing
dynamical dependence~\eqref{re} is then independent of a concrete
kind of underlying interactions.
Similarly, the local strong interactions described by the QCD
Lagrangian create hadrons which are {\it extended objects}, hence,
one may expect that some dynamical symmetries come into play.

In summary, the standpoint of the present paper is that the
observed spectral degeneracies have somewhat dynamical origin,
hence, in order to advance in understanding the spectral
degeneracies one should think in terms of internal structure of
hadrons rather than analyze dynamics and general properties of
QCD. The spirit of our work has something in common with that of
Ref.~\cite{ia2} where the spectrum-generating algebra approach was
used to deduce the $SO(4)$ dynamical symmetry from the string-like
properties of mesons.

The paper is organized as follows. In Sect.~2 we provide some
general arguments justifying the approach we will use. Sect.~3 is
devoted to construction of $O(4)$ classification for mesons. In
Sect.~4 the proposed scheme is discussed and compared with some
other approaches. We conclude in Sect.~5.

\section{Preliminary remarks}

The isospin invariance $SU(2)_f$ does not need comments, it is the
generally known vector part of the spontaneously broken chiral
$SU(2)_L\times SU(2)_R$ symmetry of the QCD Lagrangian in the
limit of vanishing current quark masses. The symmetry $I$ appears
as a consequence of the suppression of transitions between quarks
of different flavors, the so-called Zweig (or OZI) rule. The Zweig
rule is well understood in the $1/N_c$ expansion since it becomes
exact at $N_c=\infty$, i.e., in the planar limit of
QCD~\cite{hoof,hoof2}. Usually, the large-$N_c$ limit works fairly well
in the phenomenology, there are sizeable violations of the
OZI-rule only for a relatively small number of states, typically
in the scalar sector, reflecting a specific nature of those states
which results in a considerable mixture of strange and nonstrange
components. It should be noted that the $I$-symmetry is of
dynamical origin as it is not present in the QCD Lagrangian. The
symmetry $O(4)$ appears to be also dynamical, this novel symmetry
will be the subject of our discussions in what follows.

At present there are different ideas (not yet proved rigorously)
on the excited light mesons which happened to by quite successful
in a global description of the spectroscopic data. In fact, the
assumption of $O(4)$-symmetry is likely the only self-consistent
way for unification of those ideas. First of all, various
arguments and observations indicate that the spin-orbital and
spin-spin correlations are strongly suppressed in the excited
unflavored hadrons~\cite{wil,wil2,wil3,wil4,wil5,sh,glozrev}.
This suggests that,
neglecting a possible fine splitting due to such correlations and
other non-leading effects, the pattern of mass degeneracies of
mesons built from the conventional spinor quarks is the same as
that of mesons made of scalar quarks. Since the light mesons are
ultrarelativistic systems the use of the potential models is
difficult to
justify, one should rather solve the Bethe-Salpeter equation for
two scalar particles interacting through massless bosons. The
corresponding solutions reveal the $SO(4)$-degeneracy, this result
goes back to Wick and Cutkosky~\cite{wick,wick2}. The group $SO(4)$ is
known to be the dynamical degeneracy group of the nonrelativistic
hydrogen (H) atom~\cite{fock,fock2}.

The H-like $SO(4)$ degeneracy implies the dependence of
discrete spectrum on a single "principal" quantum number $n$,
\be
\label{1}
n=l+n_r+1,
\ee
where $l$ is the angular momentum and $n_r$ labels the "radial"
excitations. On the other hand, it has been observed recently~\cite{mpla,sh}
that the dependence of the meson mass $M$ on $l$ and $n_r$ indeed
enters in the combination $l+n_r$, namely, to a rather high
accuracy, the whole spectrum of excited unflavored meson resonances
can be fitted by the linear relation~\cite{mpla,prc}
\be
\label{2}
M^2=a(l+n_r)+b,
\ee
with $a\approx1.1$~GeV$^2$ and $b\approx0.7$~GeV$^2$. It is
interesting to note that the linear dependence of $M^2$ on $l+n_r$
holds in certain quasiclassical strings~\cite{baker}
(see also~\cite{npb} for the discussions based on the QCD sum rules)
and, by construction, in some AdS/QCD models~\cite{katz,katz2},
while it cannot be obtained within
the existing potential models~\cite{bicudo}, namely the
semirelativistic potential models with linearly rising potential
yield typically $M^2\sim l+cn_r$ with $c\neq1$. Thus, although
we use the nonrelativistic basis, our framework will not be
completely equivalent to old potential models. Introducing the
quark spin in the additive way as in the usual quantum mechanics,
one obtains the physical mesons with the spin $J=l,\,l\pm1$, which
possess the masses dictated by Eq.~\eqref{2}. The outlined
dynamical mechanism seems to be responsible for the emergence of
an approximate degeneracy among resonances of different spin
value. The assumption of suppression of the spin-orbital and
spin-spin correlations inside excited mesons is crucial in this
kind of reasoning, otherwise the angular momentum of $\bar{q}q$
pair and the intrinsic quark spin cannot be separated in the
relativistic systems under consideration, hence, the formulas
like Eq.~\eqref{2} may not be written.

All these arguments are quite standard, nevertheless they do not
save us from a certain uneasiness caused by the fact that we are
trying to describe the ultrarelativistic systems by means of the
unobservable nonrelativistic terms. It would be desirable to
understand deeper why the nonrelativistic basis may be useful. For
instance, consider a strong decay $A\rightarrow B+C$, where $A$,
$B$, and $C$ are some mesons. Experimentally one is able to
determine the relative angular momentum $L$ of the hadron pair $B$
and $C$. Intuitively, it is easy to imagine the following picture:
Quark and antiquark inside the hadron $A$ have the relative
momentum $l$, then the strong gluon field inside $A$ creates from
the vacuum a quark-antiquark pair, the whole system rearranges
into two colorless hadrons $B$ and $C$ which, in turn, conserve
the relative angular momentum, $L=l$, if $l_B=l_C=0$, say if $B$
and $C$ are pions. In reality, however, we should confess honestly
that we do not know and cannot imagine the internal structure of
meson $A$. But it is natural to conjecture that the observable $L$
reflects somehow this structure. A relevant example is the
observation of excited light mesons with identical quantum numbers
and very close masses, which are related to two different values
of $L$. It is reasonable to assume that their internal structure
is different, an additional argument is that these two kinds of
almost degenerate mesons always have different full width
--- this is natural as long as two different quantum systems
generically have different lifetimes. Thus, introducing $l$ and
identifying $l=L$ (plus fixing the orientation of intrinsic quark
spin) we may expect that thereby we do an unambiguous mapping of
observable $L$ onto the internal structure of observed meson,
moreover, to a certain extent we may expect that this mapping is
universal for all mesons, this permits then to establish some
relations between mesons, such as relations between masses. It
should be added also that the angular momentum $l$ and the total
quark-antiquark spin $s$ can be well defined through the
observable P- and C-parities~\cite{sh} (see their
definitions~\eqref{defPC} below). Thus, classifications in terms
of unobservable $l$ can definitely make sense, in this regard it
should be reminded that the standard $SU(3)_f$ classifications of
hadrons are also based on unobservables, which are the quarks. The
existence of noticeable mass splittings inside $l$-multiplets
should not be regarded as some kind of drawback of the proposed
scheme since they could encode an important physics (say, the spin
interactions) like the mass splittings inside the
$SU(3)_f$-multiplets.

Another argument in favour of our approach is that even
essentially relativistic models for light hadron spectrum can
possess the property that the states in their spectrum are
classified as in nonrelativistic potential models. An example of
such models is given in~\cite{berd} where the mesons are described
by a hadron string with massless spinor quarks at its ends. In
addition, it is easy to see that in the case of breaking of
classical string, a part of its angular momentum is converted into
the relative angular momentum of "splinters" and if these
"splinters" are spinless (as it usually happens in real life) this
conversion is complete due to the momentum conservation, i.e. one
has $l=L$ just as expected. As long as string models for hadrons
are known to be well motivated by QCD, our discussions above are
also well motivated.

Our approach is very different from the so-called chiral symmetry
restoration (CSR) scenario, which is claimed to be completely
relativistic and QCD-based explanation of many observed spectral
degeneracies~\cite{glozrev}. First of all, the CSR explains
degeneracies among states of equal spin, e.g., the parity
doubling, while the observed degeneracy is much
broader~\cite{epja}. A detailed comparison of our scheme with the
CSR one is presented in Sect.~4, here we would give the following
general remark. The CSR scenario treats the observed degeneracies
as a completely quantum effect, i.e., it does not have a well
understood classical limit, this point has been already criticized
in~\cite{sh} from the point of view of linearity of Regge
trajectories. The hadrons are bound states of quarks, therefore
they are described by some theory of bound states and it is quite
difficult to imagine that such a theory does not have the
quasiclassical limit. The quantum effects are decisive in
phenomena like boundary effects (e.g., the Casimir effect) or
quantum tunneling, but in bound states, they commonly result in
fine splittings of energy levels which are the next-to-leading
effects. Our standpoint is that the theory of bound quarks does
have the classical limit, we try to guess the dynamical symmetry
in this limit and use it as a starting point for further analysis.

In what follows, we proceed to explanation of observed
degeneracies in light nonstrange mesons on the base of
nonrelativistic basis, finally it will turn out that the scheme
can be reformulated in terms of observable hadron spin. The
detailed phenomenological analysis based on Eq.~\eqref{2} was
carried out in~\cite{mpla,prc} and we will not repeat it here, of
our concern will be the group-theoretical aspects and their
physical sense.

\section{Construction of $O(4)$ classification}

The light nonstrange mesons are characterized by the quantum
numbers $I^G(J^{PC})$, with the P, C, G parities defined as
\be
\label{defPC}
P=(-1)^{l+1},\quad C=(-1)^{l+s},\quad G=(-1)^{l+s+I},
\ee
where $s$ is the total quark-antiquark spin. The G-parity is not
of interest for us since it is just a combination of the
C-parity and isospin. Changing the angular momentum $l$ by one
unit we change immediately the P and C parities. Define the pure
and mixed $P$ and $C$ transformations as
\ba
\label{4}
P:& |\Delta l|=1,& l+s=const,\\
\label{5}
C:& |\Delta s|=1,& |\Delta l|=0,\\
\label{6}
PC:& |\Delta l|=1,& |\Delta s|=0.
\ea
The change of $l$ can be compensated by that of $n_r$ such that
the sum $l+n_r$ remains constant, the meson mass then is not
affected due to Eq.~\eqref{2}. The $C$-transformation preserves
the meson mass by virtue of the assumed quark spin orientation
independence of the hadron masses. Thus, there is a possibility to
relate, in some sense, the $P$ and $C$ invariances of the QCD
Lagrangian to the same invariances of the resonance spectrum.

Supplementing the $P$ and $C$ transformations defined in
Eqs.~\eqref{4}-\eqref{6} by the $I$-transformation discussed above
(the mass-conserving transitions from isosinglet channels to the isotriplet
ones and {\it vice versa}) we obtain the complete set of
transformations relating different states within a degenerate
cluster. For instance, consider the first cluster of unflavored
mesons. It is populated by the well-established states from the
PDG~\cite{pdg}, the fine splitting does not exceed 10\% of meson
mass except for the $h_1(1170)$-meson, the fine splitting is known
to reduce progressively in the higher clusters~\cite{epja}. We can
"walk" along the whole tower of states, e.g., in the following way,
\begin{multline}
a_2(1320)\xrightarrow{I}f_2(1270)\xrightarrow{CI}b_1(1235)\xrightarrow{I}h_1(1170)\xrightarrow{C}
f_1(1285)\xrightarrow{I}a_1(1260)\xrightarrow{PC}\\
\rho(1450)\xrightarrow{I}\omega(1420)\xrightarrow{CI}
\pi(1300)\xrightarrow{P}a_0(1450)\xrightarrow{I}f_0(1370)\xrightarrow{P}\eta(1295).
\label{7}
\end{multline}
Similarly, one is able to go over the resonances in the higher
towers, those clusters contain more mesons including some missing
states.

\begin{figure}
\vspace{-7cm}
\hspace{-4cm}
\includegraphics[scale=0.9]{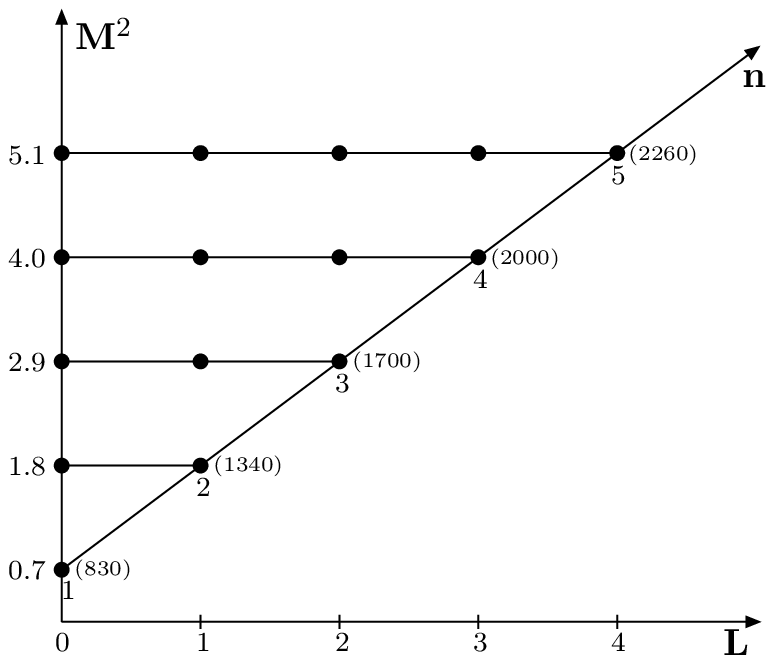}
\vspace{-16cm}
\caption{\label{f1} A graphical representation of Eq.~\eqref{2} with physical
values of parameters in GeV$^2$. The principal quantum number $n$ is defined in Eq.~\eqref{1}.
The dots denote the corresponding states (only several low-lying levels are shown).
The numbers in brackets display the predicted mean mass in MeV.}
\end{figure}

The multiplets predicted by Eq.~\eqref{2} are drawn in
Fig.~\ref{f1}. The states lying on the
diagonal line have $n_r=0$, they form the leading Regge trajectory,
with the spin being $J=l$ or $J=l+1$ in the real situations. It
should be emphasized that these resonances do not possess P-parity
doublets --- the states of equal spin and close mass but with the
opposite P-parity --- as the $P$-transformation~\eqref{4} for
such mesons cannot conserve the spin and mass simultaneously.
\begin{figure}
\vspace{-8cm}
\hspace{-5.4cm}
\includegraphics[scale=1]{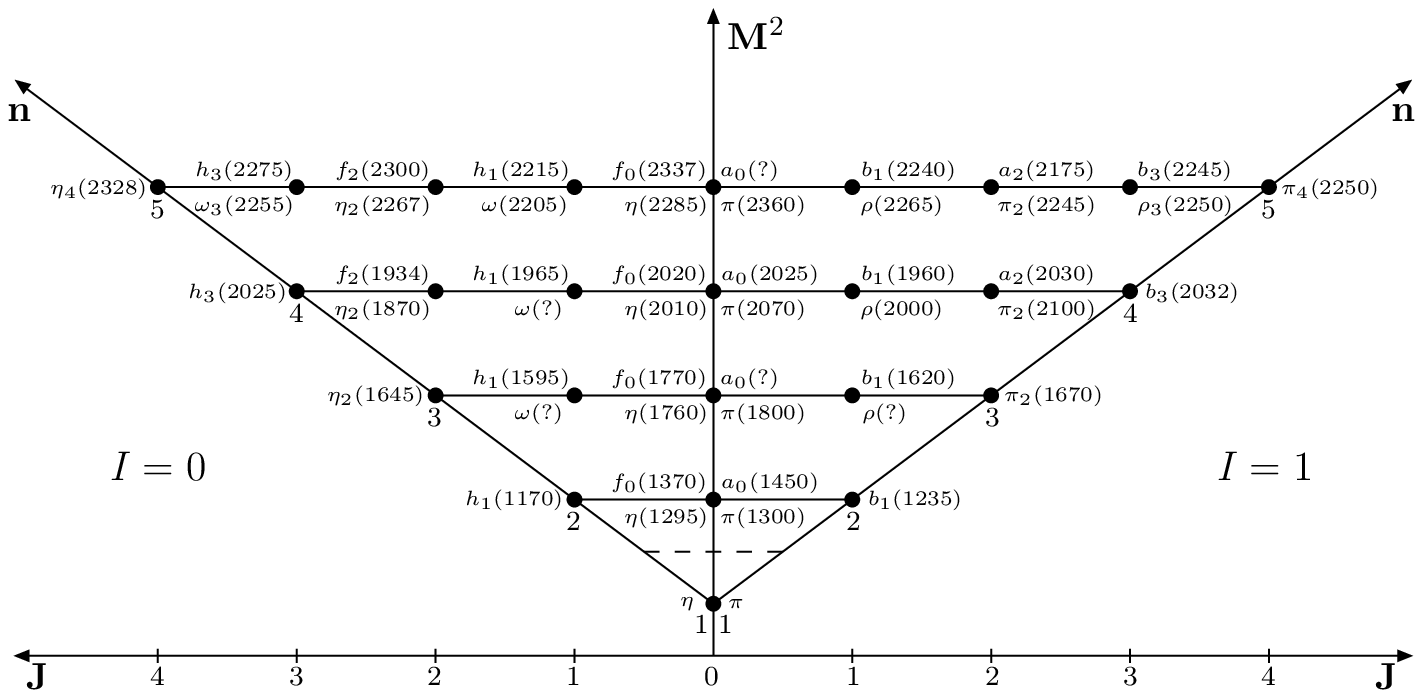}
\vspace{-17cm}
\caption{\label{f2} A hydrogen like classification for the states with $J=l$
and for their P-parity doublets. The dashed line
denotes symbolically the CSB scale, the given classification is not
expected to be reliable below this scale.}
\end{figure}
\begin{figure}
\vspace{-6cm}
\hspace{-5.4cm}
\includegraphics[scale=1]{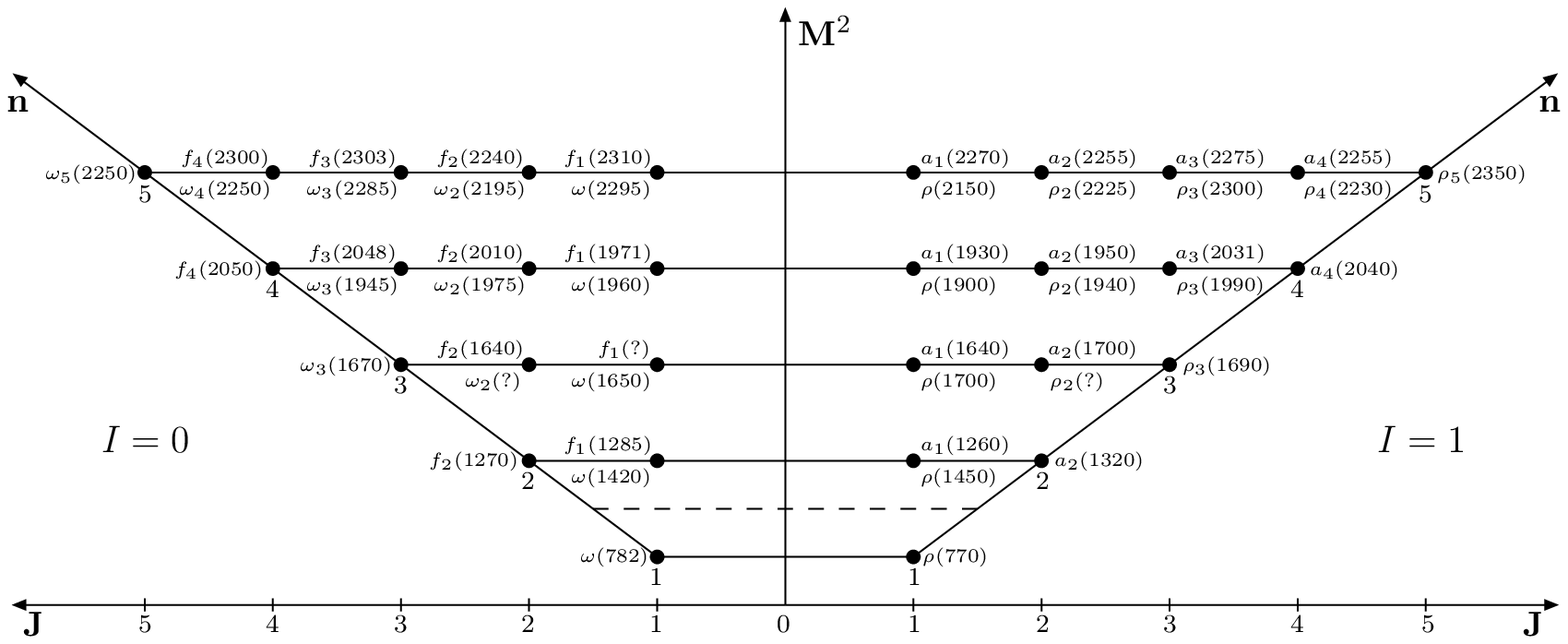}
\vspace{-18cm}
\caption{\label{f3} The same as in Fig.~\ref{f2}, but for the states having $J=l+1$
and for their PC-parity doublets.}
\end{figure}

Besides spin, a complete extension of Fig.~\ref{f1} to the real
mesons must include the isospin and doubling of both P and C
parities. The isospin can be incorporated by a reflection with
respect to the axis $M^2$, the values of mass remain intact due to
the $I$-invariance. There are two possible ways of $P$-parity
doubling, they correspond to $P$ and $PC$ transformations.
The former case is depicted in Fig.~\ref{f2}, the states on the
leading Regge trajectories have $s=0$, hence, $J=l$. The latter
possibility is displayed in Fig.~\ref{f3}, the resonances
belonging to the leading trajectories have then $s=1$, $J=l+1$. As
remarked above, the resonances on the leading trajectories are
P-parity singlets, all other states are P-parity doubled. The last
step is to superimpose Fig.~\ref{f3} on Fig.~\ref{f2} identifying
the $M^2$ axes and dashed lines and turn one of figures through
angle 90$^{\circ}$, in this way we incorporate also the C-parity.
The horizontal lines of degenerate states in Fig.~\ref{f2} and
Fig.~\ref{f3} will form then planes, the clusters of degenerate
states live on these equidistant and parallel planes. For example,
the states in cluster~\eqref{7} populate the lowest such plane. The
resulting three-dimensional picture of meson degeneracies is easy
to imagine, although the corresponding figure
appears to be beyond author's artistic abilities.

The states below 1.9~MeV in Fig.~\ref{f2} and Fig.~\ref{f3} are
taken from the PDG~\cite{pdg}, the nonstrange nature of those
resonances is usually indicated by their decay channels. Above
1.9~MeV the states are mainly from a review~\cite{bugg}, the PDG
lists them in section "Other States". A complementary test for the
non-strangeness of included states is that they belong to the
relevant families of Regge trajectories~\cite{bugg,ani,ani2,ani3}. The
numbers in brackets serve for orientation only as long as often
they refer to the traditional names of particles given by the PDG
rather than to the actual mass. For instance, the mass of
$\rho_5(2350)$ is $2330\pm35$~MeV according to the PDG~\cite{pdg}, in
the $\bar{p}p$ annihilation it is seen with the mass
$2300\pm45$~MeV~\cite{bugg}, the mass of $\omega_5(2250)$ was
estimated in the $\bar{p}p$ annihilation as
$2250\pm70$~MeV~\cite{bugg}, thus, it is not excluded that
$\rho_5(2350)$ and $\omega_5(2250)$ are exactly degenerate despite
so different numbers in brackets, which would mean the exact
$I$-symmetry for them. Another example is the $\pi_2(2100)$-meson
of the PDG, its mass looks considerably bigger than the averaged
value 2000~MeV in the corresponding cluster, however, such an
observation may turn out to be misleading since, say, in the $\bar{p}p$
annihilation this resonance was seen in the region
$2005\pm15$~MeV~\cite{bugg}. The same can be said about the
$\eta_2(1870)$-meson, which was observed in the $\bar{p}p$ annihilation
at $2030\pm16$~MeV~\cite{bugg}.
In all other cases any judgements about the fine splittings within the
clusters should be also made with caution.

Notably, within the presented classification of light nonstrange
mesons there is no place for the states $f_0(600)$, $f_0(980)$,
and $a_0(980)$, the nature of which is highly controversial.

\section{Discussions}

It is interesting to notice that although we have used
the nonrelativistic arguments in building our classification, the
final scheme turns out to be relativistic as long as formally the
spectrum depends on the spin $J$ and the number $n$ enumerating the
daughter trajectories, in principle, now one can detach from the
nonrelativistic interpretations at all, regarding Fig.~\ref{f2}
and Fig.~\ref{f3} as classifications for the states generated by
the leading Regge trajectories of unnatural, $P=(-1)^{J+1}$, and
natural, $P=(-1)^J$, P-parity, respectively.
In addition, the proposed
classification coincides with the classification of energy levels
in the relativistic H-atom, see~\cite{ijmpa} for
references. The latter scheme was used for description of the
light nonstrange baryons in 1960s (see, e.g.,~\cite{barut,barut2}; numerous
references are collected in a review~\cite{ijmpa}). In essence,
the H-like description of the light nonstrange mesons contains
only one substantial complication in comparison with the baryons
--- the resulting picture of mass degeneracies is
three-dimensional due to the existence of C-parity. The
relativistic $O(4)$ description of the H-atom emerged in
1960s from a remarkable group-theoretical discovery: The full
relativistic theory of the H-atom (without account for
electron spin) can be formulated as a dynamical group theory based on
$O(4,2)$, the conformal group. The unitary irreducible
representations of $O(4,2)$ are labelled then by $|nJm_{\pm}\rangle$,
where $m$ is the usual magnetic quantum number, $n$ is the
relativistic principal quantum number, and $\pm$ refers to the
P-parity, which is determined from the parity of the ground state.
While the
$O(4)$ symmetry relates only states within a degenerate energy
level, the $O(4,2)$ symmetry relates also different energy levels,
in our case the latter relation is given by Eq.~\eqref{2}.
One of reduction of $O(4,2)$ to $O(4)$ corresponds to the
relativistic H-atom, where all states for a given $n$ are
P-parity doublets, except the state $J=n-1$ which is a singlet.
The P-parity doubling distinguishes the relativistic $O(4)$ H-like
assignment of energy levels from the nonrelativistic $SO(4)$ one,
the group $O(4)$ is just the extension of $SO(4)$ by P-parity.
The absence of P-parity partners for the states lying on the
principal Regge trajectories is a remarkable feature of the H-based
scheme since such partners have never been observed in the mesons.

As was mentioned in Sect.~2, the most known recent explanation of
spectral degeneracies among the highly excited states is based on
the effective axial and chiral symmetry restoration at high
energies, the relevant ideas are summarized in a
review~\cite{glozrev} (see also~\cite{ijmpa,sh2}). In this regard,
it would be instructive to compare in detail our scheme with the
CSR one. Resorting to some semiclassical arguments, the latter
idea suggests that the highly excited hadrons fall into the
multiplets of approximate chiral $SU(2)_L\times SU(2)_R$ symmetry
of QCD extended by P-parity, the resulting parity-chiral group is
isomorphic to $O(4)$, we will call it $O(4)_{pc}$ in what follows.
First of all, the possible physical origins of the H-like $O(4)_H$
symmetry and that of $O(4)_{pc}$ are completely different, the
former invariance is a dynamical symmetry reflecting the internal
space structure of mesons and the centrosymmetric character of
interactions between the constituents, while the latter one is a
classical symmetry of the QCD Lagrangian. The $O(4)_H$ symmetry
can relate states with different spin, while the $O(4)_{pc}$
symmetry relates states of equal spin only.

\begin{figure}
\vspace{-4cm}
\hspace{3cm}
\includegraphics[scale=0.8]{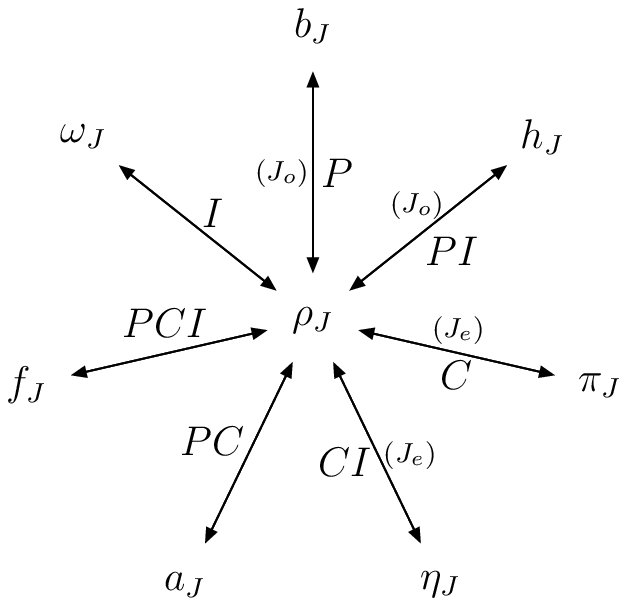}
\vspace{-15.5cm}
\caption{\label{f4} A diagram for the spin-preserving transformations
with the center $\rho_J$. The symbols $J_o$ and $J_e$ mean that
the given transformation can be performed for the odd or, respectively,
even values of spin $J$ only.}
\end{figure}

Consider as an example the $\rho_J$-mesons.
The degenerate states of equal spin value can be obtained with the
help of certain combinations of the $P$, $C$, and $I$
transformations in the way depicted in Fig.~\ref{f4}. The presented
diagram provides also all other spin-preserving transformations,
they can be trivially performed through the "center" $\rho_J$ taking
into account that double transformation of any kind is unity.
For instance, the line
$a_J\xrightarrow{PC}\rho_J\xrightarrow{P}b_J$
gives the C-parity doubling for $a_J$ and $b_J$.

The same chain of degeneracies as in Fig.~\ref{f4} follows from
a classification of mesons according to multiplets of $O(4)_{pc}$
and of axial $U(1)_A$~\cite{glozrev}.
However, there exists a crucial difference between the two
classifications: The CSR scenario predicts P-parity doublets for
all highly excited states, while the H-like scenario predicts that
the states lying on the principal Regge trajectories are P-parity
singlets, all other mesons are P-parity doubled. As mentioned above,
experimentally
P-parity doublers for the states belonging to the principal
meson Regge trajectories have not been observed, the Crystal
Barrel experiment confirmed this phenomenological
fact~\cite{mpla,epja,prc}. The absence of such P-parity partners
is a strong advantage of the H-based scheme over the CSR scenario:
Usually the states on the daughter trajectories are less reliable than
the resonances on the principal trajectories, hence, the CSR scheme fails
completely in the most reliable part of the meson spectrum.

Thus, our analysis shows that the hypothetic effective
restoration of chiral and axial symmetries of the classical
QCD Lagrangian in the upper part of the hadron spectrum does not
necessary constitute a piece of the broader degeneracy
$\mathcal{X}$ existing in that part of the spectrum (in contrast
to the point of view taken in~\cite{glozrev}). If the symmetry
$\mathcal{X}$ is of the type advocated in the present paper, the predictions
of the CSR scheme are included into $\mathcal{X}$ partly only. It is
quite important to emphasize that the "not overlapped" with $\mathcal{X}$
part of the CSR predictions lies completely in the unobserved part of the
meson spectrum.

It is interesting to mention that a pattern of P-parity
singlets similar to that of $O(4)_H$ assignment emerged
naturally in the geometrical string like and bag
like models proposed in~\cite{iachello}.

Let us try to figure out qualitatively a possible
physical origin of the $O(4)_H$ symmetry in the meson spectrum. On
the intuitive level, it is clear that both in the H-atom and in
the mesons one deals with quantum two-body systems interacting via
centrosymmetrical forces, the appearance of an universal
dynamical symmetry is then quite conceivable. In QCD, the CSB
disturbs drastically the low-energy part of the spectrum, for this
reason a manifestation of this universal dynamical symmetry should
be naturally expected above the CSB scale. The observation of the
same symmetry among the excited light baryons may indicate on
their quar-diquark structure, in fact, historically the $O(4)_H$
symmetry was first proposed for baryons on the base of analyses of
a rather rich baryon spectrum, which was available already at
1960s, see~\cite{ijmpa,barut,barut2} for references. On the other hand,
it is not excluded that the $O(4)_H$ symmetry might be given an
interpretation as a "survived" part of a broken fundamental classical
symmetry. Indeed, in the very high energy limit, the QED and QCD
Lagrangians possess the conformal invariance $O(4,2)$
(more generally, the high-energy density of states of any $d$-dimensional
renormalizable field theory is that of $O(d,2)$ conformal theory), which is
incompatible with the existence of bound states as long as the
spectrum of conformal theories is massless or continuum. From the
group-theoretical point of view, the $O(4,2)$ is also incompatible
with the existence of a finite number of degenerate states at some
energy since the unitary irreducible representations (UIR) of
$O(4,2)$ are infinite-dimensional. However, the group $O(4,2)$
contains subgroups with finite-dimensional UIR, which already are
able to accommodate the discrete spectrum and certain degeneracies
in their multiplets. The maximal such subgroup is exactly $O(4)$.
This intriguing relation of $O(4)$ and $O(4,2)$ might give a chance
to relate the observed spectral degeneracy to the fundamental theory.

\section{Conclusions}

We have proposed a classification scheme for light nonstrange
mesons which explains completely the observed approximate mass
degeneracies, only a few of states are missing and we hope they
are to be discovered in future experiments. By and large, the
accuracy of the mass degeneracies in the proposed multiplets is
similar to that of the unitary $SU(3)_f$ symmetry.
The classification looks most naturally in terms of unobservable
angular momentum of quark-antiquark pair, but it can be
reformulated also in terms of observable hadron spin.

The main message of the present work is that the observed spectrum
of light nonstrange mesons is similar to nothing but the discrete
spectrum of the classical hydrogen atom. Such ideas appeared about
forty years ago in the baryon spectroscopy, so the present
analysis may be regarded as a revival of those forgotten ideas in
application to mesons.

If this result is correct, a natural question arises as to why
the nonrelativistic symmetries can work in the excited light
hadrons, which represent ultrarelativistic systems? This question
reminds the old question why the nonrelativistic model of constituent
quarks works in the domain where naively it should not work? The
understanding of the latter problem took a long way, now we know that
clue lies somewhere in the fact that at low energies the effective
physical degrees of freedom are not those of the QCD Lagrangian, but
the exact implementation of this mechanism is still a riddle. It may
be that with the nonrelativistic symmetries in the highly excited light
hadrons we are also staying at the beginning of a long way...

In conclusion, we have tried to demonstrate that a mere observation
of hadron clusters may open the door to a new line of research,
where far-reaching results could be obtained. At present,
there exists only one experiment which systematically looked for
the excited unflavored mesons in a broad energy range, the Crystal
Barrel one~\cite{bugg}, and its results we have actively used in
this work. It would be really nice if experimentalists taught
us more about the particle content of the hadron clusters,
elevating thereby the clustering from the present somewhat speculative
level to a rather unexpected new direction in the particle physics.
A tentative program of relevant physical experiments is proposed
in~\cite{bugg2}.

\section*{Acknowledgments}

The work was supported by the Ministry of Education of
Russian Federation, grant RNP.2.1.1.1112, by the Government
of Sankt-Petersburg, grant 29-04/23, and by the Alexander von
Humboldt Foundation.

\end{document}